\documentclass[useAMS,usenatbib]{mn2e}
\usepackage{rotating}

\newcommand{\unit}[1]{\,{\rm #1}}
\renewcommand{\d}{{\,\rm d}}

\title[Slowly increasing elongations of non-spherical asteroids caused by
  collisions]{Slowly increasing elongations of non-spherical asteroids caused by collisions}
\author[Tom\'{a}\v{s} Henych, Petr Pravec]{Tom\'{a}\v{s} Henych$^{1}$\thanks{E-mail:ftom@physics.muni.cz (TH)}, Petr Pravec$^{1}$
\\
$^{1}$Astronomical Institute, The Czech Academy of Sciences, Fri\v{c}ova 298, CZ-25165 Ond\v{r}ejov, Czech Republic}

\begin{document}
\date{Accepted 2015 September 11 \hspace*{3em} Received 2015 September 10; in
  original form 2015 August 03}

\pagerange{\pageref{firstpage}--\pageref{lastpage}} \pubyear{2015}

\maketitle

\label{firstpage}

\begin{abstract}
Asteroids are frequently colliding with small projectiles. Although
each individual small collision is not very important, their
cumulative effect can substantially change topography and also the
overall shape of an asteroid. We run simulations of random
collisions onto a single target asteroid represented by triaxial ellipsoid. We
investigated asteroids of several hundred meters to about
$18\unit{km}$ in diameter for which we assumed all material excavated
by the collision to escape the asteroid. The cumulative effect of
these collisions is an increasing elongation of the asteroid
figure. However, the estimated timescale of this process is much
longer than the collisional lifetime of asteroids. Therefore, we
conclude that small collisions are probably not responsible for the
overall shape of small asteroids.
\end{abstract}

\begin{keywords}
minor planets, asteroids: general -- methods: numerical.
\end{keywords}

\section{Introduction}\label{intro}
Collisions between asteroids are a very important process, which affects
their spins (\citealt{harris1979a}, \citealt{harris1979b}), size
distribution (\citealt{dohnanyi1969}, \citealt{cellino1991}) and also
their shapes (\citealt{catullo1984}, \citealt{farinella1997},
\citealt{tanga2009}). Subcatastrophic collisions, however, are not
very well described yet and in evolutionary models of asteroid
populations, they are only roughly approximated. 

The role of small collisions on the overall shape of asteroids was
studied in two-dimensional model of \cite{ronca1979}. They used
similar assumptions in the construction of their model as we do and
for the initially elliptical asteroid model they obtained increasing
elongation of its shape. Later, \cite{harris1990} mentioned this
effect when discussing collisional evolution of the spin of
nonspherical asteroids. 

Numerical investigation of asteroid shape evolution by collisions was
presented by \cite{korycansky2001} and \cite{korycansky2003}
(hereafter KA03). Their model is based on excavation, orbit and
reimpact of the ejecta from the cratering event and also on the
conservation of the maximum angle of repose of the material on the
surface of the model asteroid. KA03 conclude that the final shape is
mostly oblate ellipsoid indepedently of the initial conditions. They also
calculated the timescale of such impact-induced reshaping of the
asteroid to be more than an order of magnitude longer than
catastrophic disruption timescale.

\cite{szabo2008} report about shape distribution of asteroid families
derived from multi-epoch photometry of the Sloan Digital Sky Survey
Moving Object Catalog. They conclude that younger families seem to
contain more elongated asteroids, while older families contain more
rounded asteroids. In their opinion, this is in agreement with the
predictions of an impact-driven evolution of the asteroid shapes
(KA03). \cite{domokos2009} studied the role of micro-collisions on the
overall shape of asteroids in their averaged continuum abrasion
model. Their asteroid models evolve to prolonged shapes that possess
large, flat areas separated by edges. 

In the present paper we want to discuss the effect of a large number of
subcatastrophic collisions on the shape of an asteroid. This problem
caught our interest when constructing an evolutionary model of
asteroid rotations influenced by subcatastrophic collisions 
focused on tumbling asteroids (\citealt{pravec2005}). We noticed that
our model asteroids subjected to collisions became ever more elongated
than at the beginning of the simulation. Since we wanted to ensure it
was not just a numerical issue of our model, we started to investigate
this effect.

We give a detailed description of our model and the method used in
Section~\ref{model}. Section~\ref{timescale} contains an estimate of
timescale of the increasing elongation process. Numerical results are
given in Section~\ref{results}. In Section~\ref{disc}, we discuss
some implications and caveats of the present research and conclusions
are given in Section~\ref{conc}.

\section{Increasing elongation of the model asteroid}\label{model}
An obvious effect of a small collision is a formation of impact crater
on the surface of an asteroid. There is, however, a cumulative effect
of a large number of such cratering events as well. In our analytical
model of subcatastrophic collisions between a target asteroid and a large
number of small projectiles, we noticed that the shape of an initially
elongated asteroid becomes gradually more elongated. More exactly,
it was changing the shape of a dynamically equivalent equal-mass
ellipsoid. Its semiaxes are defined as (\citealt{kaasalainen2001},
\citealt{pravec2014})
\begin{eqnarray}
a_{\rm dyn}^2 = \frac{I_b-I_a+1}{I_a+I_b-1}\qquad b_{\rm dyn}^2 =
\frac{I_a-I_b+1}{I_a+I_b-1}\qquad c_{\rm dyn} = 1\,,\nonumber
\end{eqnarray}
and
\begin{eqnarray}
I_a = I_1/I_3\qquad I_b = I_2/I_3\,,
\end{eqnarray}
where $I_1$, $I_2$ and $I_3$ are the principal moments of inertia of the
target asteroid and $I_a$ and $I_b$ are normalized principal moments of inertia
of the dynamically equivalent ellipsoid; $I_c=1$.

In our model, the target asteroid was represented by a triaxial
homogeneous ellipsoid. We ran a series of consecutive collisions with
the target, each forming an impact crater on the surface of the ellipsoid 
according to the scaling laws of \cite{holsapple1993} and
\cite{holsapple_housen2007}, with the angular momentum transfer
efficiency based on laboratory experiments of \cite{yanagisawa1996}
and \cite{yanagisawa2000} and updating the inertia tensor of the target
asteroid after every single collision.

The impact speed was $5\unit{km\,s^{-1}}$, which is the median impact
speed in the central regions of the Main Asteroid Belt
(\citealt{bottke2005}). The geometry of collisions was isotropic with
respect to the target surface (the probability of a collision from any
direction was the same) with one exception: We only allowed collisions
for which incidence angle (the angle between the impact velocity
vector and the normal to the surface at the impact site) was less 
than 75 degrees. We simply wanted to avoid almost tangent impact
geometries, since the formation of impact craters for such impact
geometries is not very well understood yet.

The target was in a principal axis rotation state before the
collision. Initial rotation period was chosen randomly according to
the observed distribution of rotation periods of small Main Belt/Mars
Crossing asteroids (\citealt{pravec2008}, updated on 2014-04-20). Size
of the projectile was random according to a power-law distribution 
(\citealt{bottke2005}). 

After every collision, rotation state and shape of the target
asteroid changed. However, in this paper we only focus on a change
of the shape and we quantify this change as follows. The collision
formed an impact crater on the surface of the ellipsoid. We therefore
calculated an inertia tensor of the mass that fills the crater and
subtracted it from the inertia tensor of the ellipsoid. Then, we updated
lengths of axes of the ellipsoid using the values of the dynamically
equivalent ellipsoid. We also reset the shape of the target asteroid to
the triaxial ellipsoid (erased the impact crater) keeping its density
on the initial value (which lead to decreasing size of the
target asteroid). 

In our simulations, we only investigated the effect of subcatastrophic
collisions. This corresponds to the calm period of an asteroid's life
between larger shattering or even dispersal events (shattering causes
large scale damage to the asteroid, but most of its pieces remain
gravitationally bound; dispersal event reduces the target to one-half
its original mass, \citealt{melosh1997}, \citealt{housen2009}). We used
two criteria to recognize a subcatastrophic collision. First, we
compared the specific energy of the collision with the shattering
criterion. In our calculations, we followed \cite{stewart2012}, who
derived the dispersal criterion in gravity regime of collisions in
their numerical simulations. We calculated the shattering criterion as
1/4 of their value according to \cite{housen2009}. Second, we
calculated the maximum size of an impact crater for a given size of
the target asteroid according to the relation of
\cite{burchell2010}. The collision was subcatastrophic if both these
conditions were met.

There is a set of mechanical properties of the target and
the projectile, which can be varied in a physically plausible
range. We did this to some extent in our previous paper, 
\cite{henych2013}, and we found that these values affect the size of 
an impact crater formed on the surface of the target
asteroid. Therefore, they affect the timescale of increasing
elongation of the target's shape. In the present research, we have
chosen some plausible values of those parameters and held it constant
throughout our simulations as specified in
Table~\ref{const_values}. The density of the target and of projectiles 
was the same, $2\unit{g\,cm^{-3}}$.

\begin{table*}
\begin{minipage}{75mm}
 \caption{Scaling and material constants for non-porous target. $K_1$
 and $K_2$ are experiment-based coefficients, $\mu$ and $\nu$ are the
 scaling parameters, $\bar{Y}$ is a material strength and $K_r$ and
 $K_d$ are the shape constants, cf.~\protect\cite{holsapple-web}.}
 \label{const_values}
 \begin{tabular}{c c c c c c c}
 \hline
 $K_1$ & $K_2$ & $\mu$ & $\nu$ & $\bar{Y}\,(\rm{MPa})$ & $K_r$ & $K_d$ \\
 \hline
 0.095 & 0.257 & 0.55 & 0.33 & 1 & 1.1 & 0.6\\
 \hline
 \end{tabular}
\end{minipage}
\end{table*}

Since we are mostly interested in the small asteroid population (bodies of
several kilometers in diameter), we ran our simulations for asteroids
of mean diameters $0.6$--$17.6\unit{km}$ (see
Table~\ref{targets_projectiles}). The upper limit was
determined by the fact that almost all material excavated by the
hypervelocity impact of a projectile is ejected from the asteroid for
asteroid diameters of less than about $18$--$20\unit{km}$ according to
scaling laws (\citealt{housen2011}). This was
calculated by using the scaling laws for ejecta speeds as described by
 \cite{housen2011} and comparing it with the surface escape speed.    

The mean diameter of the target is
\begin{equation}
  D_{\rm t}=2(abc)^{1/3}\,,
\end{equation}
where $a$, $b$ and $c$ ($a\geq b\geq c$) are the semiaxes of the
triaxial ellipsoid representing the target asteroid. In our
simulations, we concentrated on two effects\=the shape and the size
dependence of the collisional erosion. Therefore, we ran longer series
of $1{,}000$ collisions for various target asteroid sizes (see
Table~\ref{targets_projectiles}) and fixed initial axial ratios of
$a/c=2.0$ and $b/c=1.4$. For each target size we ran three separate
runs to exclude any numerical artifacts. Then, we ran shorter series
of about $110$ consecutive collisions, where we fixed $c=0.5\unit{km}$
and varied the initial semiaxes ratios (see Table~\ref{semiaxes}).

\begin{table*}
\begin{minipage}{80mm}
 \caption{Effective diameters of model asteroids in long series of
   collisions and limiting projectile sizes. The fixed initial semiaxes
   ratios of the target were $a/c=2.0$ and $b/c=1.4$, $D_{\rm t}$
   is the target mean diameter, $D_{\rm p,max}$ is the largest
   projectile, $D_{\rm p,min}$ is the smallest one.}
 \label{targets_projectiles}
 \begin{tabular}{l c c c c c c c}
 \hline
 $c~[\rm{km}]$ & 0.2 & 0.35 & 0.5 & 1.5 & 2.5 & 4.25 & 6.25 \\
 $D_{\rm t}~[\rm{km}]$ & 0.6 & 1.0 & 1.4 & 4.2 & 7.0 & 12.0 & 17.6 \\
 $D_{\rm p,max}~[\rm{m}]$ & 5.5 & 10.5 & 16.0 & 78.0 & 150.0 & 291.0 & 437.0 \\
 $D_{\rm p,min}~[\rm{m}]$ & 1.0 & 1.0 & 1.0 & 1.0 & 1.0 & 1.0 & 1.0 \\
 \hline
 \end{tabular}
\end{minipage}
\end{table*}

\begin{table*}
\begin{minipage}{85mm}
 \caption{Semiaxes ratios of the ellipsoid representing the target
   asteroid used in our simulations (semiaxes $a$, $b$, $c$). The
   shortest axis was $c=0.5\unit{km}$.}
 \label{semiaxes}
 \begin{tabular}{c c c c c c c c c c c}
 \hline
 $a/c$ & 1.1 & 1.2 & 1.3 & 1.3 & 1.30 & 1.4 & 1.4 & 1.4 & 1.5 & 2.0 \\
 $b/c$ & 1.1 & 1.2 & 1.3 & 1.2 & 1.14 & 1.4 & 1.1 & 1.2 & 1.2 & 1.4 \\
 \hline
 \end{tabular}
\end{minipage}
\end{table*}

\section{Timescale estimates}\label{timescale}
The importance of erosional process for evolution of asteroid shapes
is determined by the timescale of this process. We calculated the
timescale in two different ways and obtained very different
results. We also have the timescale indicated by the numerical
simulations as explained in a following subsection. We discuss the
differences between the two timescale calculations at the
end of this section after their description. 

\subsection{Timescale of the erosion from the simulations}
The timescale of the increasing elongation of the asteroid figure can
be estimated directly from the simulations. We filtered size
of the projectiles for a specific size of the target to keep the
collisions subcatastrophic (see
Table~\ref{targets_projectiles}). However, we also saved information
about discarded projectiles, whether too small or too
large. Therefore, we have the relative timescale of the increasing
elongation in comparison to the catastrophic disruption of the target
asteroid caused by projectiles larger than upper size limit.

In all our simulation runs with $1{,}000$ subcatastrophic collisions,
the target asteroid was hit be the projectile larger than upper size
limit after a few subcatastrophic collisions. Therefore, we estimate
the timescale of the process to be at least two orders of magnitude
longer than the collisional lifetime of the target asteroid. 

\subsection{Ejected mass measure}\label{ej_mass_mesure}
The first and probably more reliable and solid calculation compares
the erosional effect of projectiles in terms of excavated and ejected
mass. As noted above, practically all mass excavated by the projectile
escapes the target asteroid for sizes used in our calculations. The
total eroded mass by the population of projectiles can be calculated
as 
\begin{equation}
 M_{\rm er}=\int^{D_{\rm p,max}}_{D_{\rm p,min}}\frac{1}{4}p_{\rm i}(D_{\rm t}+D_{\rm p})^2\Delta t AV_{\rm p}n(D_{\rm p})\d{}D_{\rm p}\,,
 \label{eroded_mass}
\end{equation}
where $p_{\rm i}$ is the intrinsic probability of collision, $D_{\rm
  t}$ and $D_{\rm p}$ are the target and the projectile diameters,
respectively, $\Delta t$ is the timescale we are looking for, $V_{\rm
  p}$ is the volume of the projectile, $A$ is a factor describing
the excavated material volume, $n(D_{\rm p})$ is the distribution
function of the projectile sizes and $D_{\rm p,min}$ and $D_{\rm
  p,max}$ are the minimum and maximum projectile size, respectively. 

\begin{figure}
        \centering
\includegraphics[angle=270,width=\columnwidth]{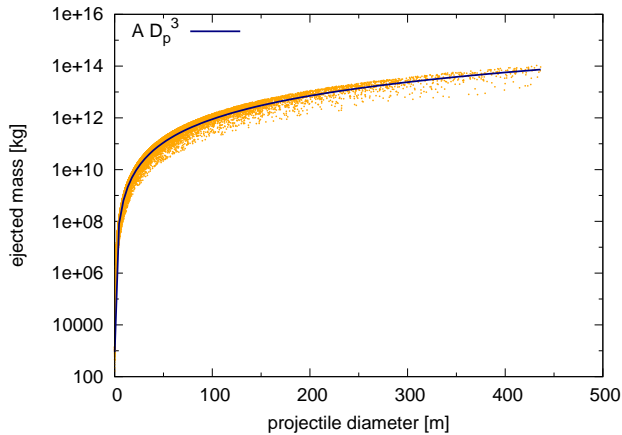}
\caption{Erosion efficiency of projectiles in all our simulations
  measured by ejected mass vs. projectile diameter. The ejected mass
  is proportional to the volume of the projectile. The vertical spread
of the points for a specific projectile diameter is caused by various
impact geometries.}
\label{erosion_eff}
\end{figure}

We define the timescale of erosion process as a time interval over
which half the target initial mass is excavated by subcatastrophic
collisions. Then, $M_{\rm er}=0.5M_{\rm t,initial}$ and when we
integrate Eqn.~\ref{eroded_mass} we obtain the timescale
\begin{equation}
 \Delta t = \frac{12(s+4)M_{\rm t,initial}}{p_{\rm i}\upi ABD^2_{\rm
     t}(D_{\rm p,max\phantom{i}}^{s+4}\!\!-D_{\rm p,min}^{s+4})}\,,
 \label{timescale_mass}
\end{equation}
where we used $V_{\rm p}=D^3_{\rm p}\upi/6$ for spherical projectiles,
$s=-2.574$ is the exponent of the distribution $n(D_{\rm p})=B D_{\rm
  p}^s$, $D_{\rm p}$ is the projectile  diameter and we approximated
$(D_{\rm t}+D_{\rm p})^2\approx D^2_{\rm t}$, since the projectile's
diameter is much smaller than the target's diameter (few percent for the
largest projectiles in our simulations). For $s=-2.574$ used in our
simulations, the calculation of $\Delta t$ is not very sensitive to
specific values of $D_{\rm p,max}$ and $D_{\rm p,min}$, because we
measure the erosion efficiency of every projectile that collided with
the target asteroid. When we set $D_{\rm p,min}=1\unit{mm}$ instead of
$1\unit{m}$ used in our simulations, the calculated timescale barely
changes. The constant $A$ can be derived from the scaling laws or
found from the graph of ejected mass vs. projectile diameter of all
our simulations (see Fig.~\ref{erosion_eff}). The constant
$B=5.07\unit{km^{-s-1}}$ for the projectile size distribution we used
(\citealt{bottke2005}). More details on derivation of these relations
are in Appendix~\ref{app_time_mass}.

\subsection{Number of collisions measure}\label{num_coll_measure}
The second approach is based on a calculation of the number of collisions
with specified population of projectiles for a single target. 

We first calculate the collisional lifetime of a model asteroid as 
(\citealt{farinella1998}) 
\begin{equation}
 t_{\rm shatter}=20\unit{Myr}\sqrt{R_{\rm t}}\,,
 \label{tau}
\end{equation}
where $R_{\rm t}$ is the mean target asteroid radius in meters. We
note that this collisional lifetime estimate is derived for smaller
asteroids and it underestimates lifetime for asteroids in our
simulations. The collisional lifetime for asteroids larger than
several hundred meters as calculated by \cite{bottke2005} is more
complicated function, but in general it has a steeper slope than that of
\cite{farinella1998}. 

Then, we calculate a number of subcatastrophic collisions that the
asteroid experiences during its collisional lifetime. In our
simulations, we introduced a cutoff of the projectile size equal to
$1\unit{m}$ because of quickly growing calculation expenses with
decreasing size of projectiles. We think that smaller projectiles are
not very relevant for this effect but more extensive calculations will
have to be done to check this in the future.

The number of collisions for a single target is (\citealt{obrien2005})
\begin{equation}
 n_{\rm coll}=p_{\rm i}(R_{\rm t}+R_{\rm p})^2N_{\rm p}\Delta t\,,
 \label{ncoll}
\end{equation}
where $p_{\rm i}$ is the intrinsic probability of collision, $R_{\rm
  t}$ and $R_{\rm p}$ are the target and the projectile radii,
respectively, $N_{\rm p}$ is the number of projectiles in 
this region and $\Delta t$ is the time interval for 
which we calculate the number of collisions. We take this equal to the 
$t_{\rm shatter}$ as calculated by Eqn.~\ref{tau}. We assume the model
asteroid is a member of the Main Asteroid Belt. The intrinsic
probability is taken from \cite{cibulkova2014} as $p_{\rm
  i}=4.91\times10^{-18}\unit{km^{-2}yr^{-1}}$.

The number of projectiles is calculated assuming that the
distribution of asteroid sizes in the Main Asteroid Belt is a power law as
already mentioned above (\citealt{bottke2005}). The number of bodies
in the size interval from $D_{\rm p,min}$ to $D_{\rm p,max}$ is
\begin{equation}
 N(D_{\rm p,min},D_{\rm p,max})=\frac{10^q}{s+1}(D_{\rm p,max}^{s+1}-D_{\rm
   p,min}^{s+1})\,,
 \label{pop}
\end{equation}
where we use the same size distribution as in
Eqns.~\ref{eroded_mass} and~\ref{timescale_mass} and $q={\rm log}_{10}
B$. The projectile sizes used for various target sizes are in
Table~\ref{targets_projectiles}. The upper limit is given by
disruption criterion (\citealt{stewart2012} and \citealt{housen2009})
and the lower limit is chosen somewhat arbitrarily to keep the
computational expenses reasonable. Finally, a rough estimate of a
timescale of the increasing elongation can be calculated. It is 
\begin{equation}
 t_{\rm reshape}=\frac{t_{\rm shatter}}{n_{\rm
     coll}}\frac{\delta_{\rm reshape}}{\delta_{\rm avg}}\,,
 \label{reshape}
\end{equation}
where $\delta_{\rm reshape}$ is the magnitude of a relative change of
any semiaxis of the asteroid after that the asteroid will experience
some abrupt change of shape and $\delta_{\rm avg}$ is the average relative
change of the same ratio per one collision. We discuss the former value
in the next paragraph, the latter value comes from our 
simulations. Also note, that this timescale actually does not depend 
on the asteroid lifetime (put the Eqn.~\ref{ncoll} to the
Eqn.~\ref{reshape}).

According to \cite{harris2009}, there are no photometric observations
of asteroids implying bodily axial ratio higher than about
$3$. Therefore, we chose the value of $\delta_{\rm reshape}=3/2$ for
the most elongated initial asteroid model used in our simulations
(initial axial ratios $2\!:\!1.4\!:\!1$) and $\delta_{\rm
  reshape}\simeq2.7$ for an almost spherical initial asteroid model
(initial axial ratios $1.1\!:\!1.1\!:\!1$).

\subsection{Comparison of the two measures}\label{comp}
We decided to prefer the timescale measure based on the ejected mass
which is a kind of efficiency measure for every projectile. This
measure gives more realistic timescale estimate and is also consistent
with the relative timescale of the process based on our
simulations. 

When we analyzed our simulations we concluded that substantial change
of the axial ratio was only caused by the very largest projectiles (or
the collisions close to the shattering threshold). Even though smaller
projectiles are more populous thanks to the power-law distribution of
their sizes (the exponent being lower than $-2$), their cumulative
effect does not balance the effect of the largest
projectiles. It is necessary to somehow weight the excavation
efficiency of the projectiles otherwise the $n_{\rm coll}$ is
overestimated and $t_{\rm shatter}$ is consequently
underestimated. Therefore the timescale measure based solely on the
number of all collisions is unrealistically skewed to the small values.

\section{Results}\label{results}
In our simulations, we observed that the elongation of the target
asteroid gradually increases with increasing number of
collisions. As described in Section~\ref{model}, we did not calculate
a change of the actual shape of the target but rather a change of the
shape of a dynamically equivalent equal mass ellipsoid. We calculated
its semiaxes from the inertia tensor of the target asteroid gradually
changed by consecutive collisions forming craters and hence removing
material from the asteroid.  

\begin{figure}
        \centering
\includegraphics[angle=270,width=\columnwidth]{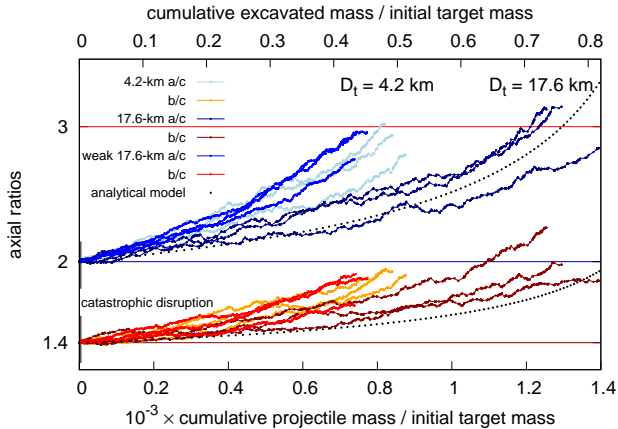}
\caption{Series of successive collisions onto two asteroids of
  various mean diameters represented by ellipsoids with initial
  $a/c=2.0$ and $b/c=1.4$. The lower abscissa is cumulative projectile
  mass divided by the initial target mass, the upper abscissa is
  cumulative excavated mass divided by the initial target
  mass. Changing axial ratios are in the ordinate. We also plot the
  simple analytical model of {\protect\cite{harris1990}} (black dots),
  the ``weak 17.6-km'' asteroid model with the same shattering
  threshold as the 4.2-km asteroid model and we mark the collisional
  lifetime of the model asteroids. See text for details.}
\label{graph_ax_rat_evo}
\end{figure}

An example of three series of collisions onto the target asteroid with
initial semiaxes ratios of $a/c=2.0$ and $b/c=1.4$ and mean diameters
of $4.2$ and $17.6\unit{km}$, respectively, is shown in
Fig.~\ref{graph_ax_rat_evo}. The graph shows the evolution of $a/c$,
and $b/c$ axial ratios vs. cumulative projectile mass divided by the
initial target mass on the lower abscissa and cumulative excavated
mass divided by the initial target mass on the upper abscissa. It can
be seen that the two ratios tend to increase as more projectiles
impact the target. Other simulations look similar to this example.

In Fig.~\ref{graph_ax_rat_evo}, we also plot three series of 
collisions for a model of a ``weak 17.6-km'' asteroid for which we set 
the same shattering threshold $Q^*_{\rm S}$ as for the 4.2-km 
asteroid. We ran these series to evaluate our hypothesis that the
apparently faster erosion of the larger asteroid models is an artifact
caused by their higher shattering thresholds as will be discussed in
Sect.~\ref{disc}. The black dots in the graph depict the analytical
model of \cite{harris1990} which represents the erosion by removal of
the constant-depth layer from the ellipsoidal asteroid model. This
model is size-independent (contrary to our model) and its apparent
consistency with collisional erosion of our 17.6-km asteroid model is
largely coincidental.

In the following Fig.~\ref{timescale_graph}, there are results of the 
timescale calculation of the erosional process for various size of the target
asteroid. It shows both, the time needed to erode half the initial
target mass by subcatastrophic collisions off the asteroid
(Sect.~\ref{ej_mass_mesure}) and also the time of reshaping the target
asteroid from the initial axial ratio $a/c=2.0$ to the axial ratio
$a/c=3.0$ (Sect.~\ref{num_coll_measure}). The timescale is in
millions of years. It is obvious from Fig.~\ref{timescale_graph}
that both timescales are much longer than collisional lifetime for most 
asteroids in our simulations.

The displayed timescales calculated by two different methods are
similar for smaller asteroids but they diverge for larger
asteroids. As already discussed in Sect.~\ref{comp}., we prefer the
timescale calculation based on the ejected mass (orange circles in
Fig.~\ref{timescale_graph}) and all our conclusions come from this
calculation.

\begin{figure}
        \centering
\includegraphics[angle=270,width=\columnwidth]{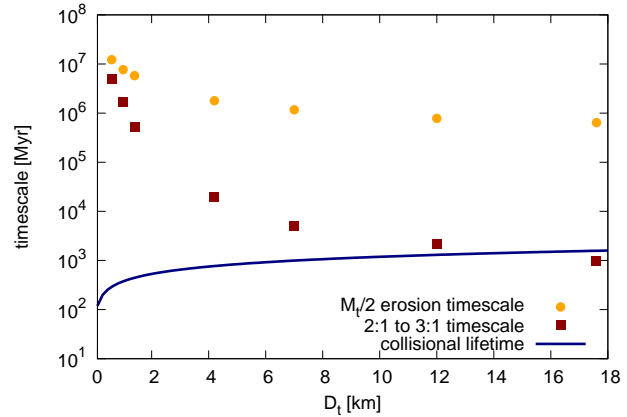}
\caption{The timescale of the collisional erosion calculated in two
  ways. Orange circles are calculated as the time
  after which half the initial asteroid mass is ejected by
  collisions, dark-red squares are calculated using the number of
  small collisions experienced by the target asteroid during its
  collisional lifetime. The collisional lifetime is calculated by
  Eqn.~\ref{tau} (\protect\citealt{farinella1998}).}
\label{timescale_graph}
\end{figure}

The initial elongation of the asteroid (shorter series of collisions)
leads to faster erosion. The timescale calculation is affected by
larger mean diameter of an asteroid $D_{\rm t}$ in
Eqn.~\ref{timescale_mass} and also by wider interval of projectile
sizes for more elongated asteroids.

\section{Discussion}\label{disc}
The explanation for the results of our simulations can be found in
\cite{harris1990}. Cratering erosion tends to change all dimensions by the
same amount on average. Imagine a triaxial ellipsoid subjected to
isotropic random impacts. We will concentrate only on the changing
dimensions of the longest and the shortest axes of the ellipsoid for a
while. We observe that impacts erode relatively larger part of the
shortest axis of the ellipsoid than its longest axis. This
leads to the increasing elongation of the ellipsoid or the ratio of
the longest to the shortest axes leghts. That is exactly what we see
in our present simulations as shown in Fig.~\ref{graph_ax_rat_evo}. 

However, from the results it is obvious that the timescale of this
process is several orders of magnitude longer than the collisional
lifetime of the asteroid. It is also visible in
Fig.~\ref{timescale_graph} that for increasing size of the target
asteroid the erosion is faster. It is probably caused by the
increasing specific energy threshold in gravity regime 
for the asteroid sizes from about $0.6\unit{km}$ used in our
simulations (see, e.g., a review by~\citealt{holsapple2002} and their
Fig.~6). It means that larger asteroid can sustain larger collisions
which excavate more material. In Fig~\ref{graph_ax_rat_evo}, it
can be seen that the ``weak 17.6-km'' asteroid model behaves much
like the 4.2-km asteroid because it cannot sustain larger craters
than this smaller asteroid. Therefore, we think that this
strengthening (by gravity) of asteroids with increasing size is
responsible for faster collisional erosion.

\begin{figure}
        \centering
\includegraphics[angle=270,width=\columnwidth]{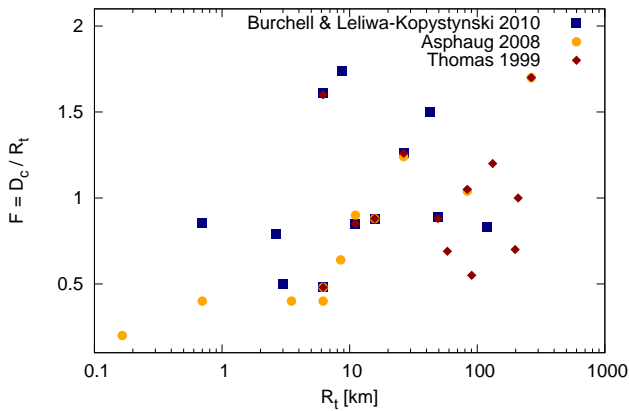}
\caption{Relative sizes of large craters observed on small solar
  system bodies vs. mean body radius. Relative size $F$ is a ratio of
  the crater diameter and the mean body radius. Larger bodies usually
  bear relatively larger impact craters. Overlapping points are the
  same craters on the same objects.}
\label{graph_crat_small_bodies}
\end{figure}

This is also supported by direct observations of large impact craters
on various small solar system bodies. We illustrate this in
Fig.~\ref{graph_crat_small_bodies} which we compiled from crater sizes
and the parent bodies sizes listed in \cite{thomas1999},
\cite{asphaug2008} and \cite{burchell2010}. The graph shows that
larger bodies bear relatively larger impact craters.

We came to the similar conclusion as KA03, however, our particular
results differ substantially. Our models are quite different and we
think it could be useful to combine their main features to study the
role of small collisions in a more complex way.

In our model, we did not work with the actual shape of the asteroid
but with the dynamically equivalent ellipsoid as described
above. Probably more serious drawback of our model is the absence of 
``landsliding'' that was present in KA03 and is probably important for
a shape relaxation to equilibrium (there is a limit of the angle
of repose of the surface material observed on the asteroids studied by
spacecraft mission as well as predicted by theory). 

On the other hand, our model deals with angular momentum evolution
of the target asteroid caused by projectiles, mass loss caused by
impact excavation and we also allow for large craters up to the
shattering threshold. All these characteristics were mentioned by KA03 as
future plans of the modelling of impact evolution of shapes of
asteroids. 

Important difference between our and KA03's model is that we do not
model ejecta's orbits and their reimpact on the surface of the target
asteroid. In our model, all ejected material escapes the
asteroid. This makes the model only valid for small asteroids in the
size range of several hundred meters to about $18$--$20\unit{km}$
(see Sect.~\ref{model} for details).  

We also differ in the resulting shape of the asteroids subjected to a
series of small collisions. While KA03 obtained oblate shape in most
of their simulations, we predict the elongation will grow for an
already elongated initial shape. Our result seems to be consistent
with an expected behaviour of large number of collisions as noted
above. KA03 also noted the dependence of the result on the initial
spin rate. We cannot verify this result since our model asteroids had
random rotation period as described in Sect.~\ref{model}.  

We predict even longer timescale of the erosion than KA03 which can be
attributed to different construction of the model and the 
timescale calculation. However, the qualitative result is the same\=the
timescale is several orders of magnitude longer than the collisional
lifetime of the asteroid for the projectile population exponent $s=-2.574$.

What follows are few caveats concerning the construction of our
model. It has several points that are affected by large
uncertainties. First, scaling laws for impact crater formation were
developed for halfspace impacts, not for impacts onto curved surfaces
and also we used earth rock material constants which may or may not be
realistic.  

Second, the catastrophic disruption threshold comes from
numerical simulations and scaling and is only one of many that can be
found in literature as there is quite a large uncertainty in this.

Third, there is a large uncertainty in population estimates of such
small projectiles (and also targets), since we are well beyond present
observation size limit. The power-law exponent is taken from the
numerical model of the Main Asteroid Belt evolution of
\cite{bottke2005}. Other values are possible and would lead to faster
or slower evolution of the asteroid shape.

Fourth, there is quite a questionable assumption about the isotropic
distribution of projectiles. We assumed that the Main Belt asteroids
have some moderate range in orbit inclination and also that the
rotation axes of the target asteroids are randomly oriented. This
last assumption may be erroneous, though, because small Main Belt
asteroids may be trapped in Slivan state (\citealt{vokrouhlicky2003})
caused by the Yarkovsky--O'Keefe--Radzievskii--Paddack (YORP) effect
which makes obliquity of their rotation axis gain specific values. It
is, however, not clear whether this mechanism also works for smaller,
km-sized asteroids. The above mentioned paper indicates that asteroids
smaller than about $10\unit{km}$ may be able to leave the resonant spin
state.

\section{Conclusions}\label{conc}
In a present paper we describe the effect of increasing elongation of
a model asteroid represented by a homogeneous triaxial ellipsoid
caused by consecutive small collisions. The changing shape of our model
asteroid was indicated by changing intertia tensor and axial ratios of
a dynamically equivalent equal mass ellipsoid. The effect can be
explained by the fact that the shorter axis of the ellipsoid decreases
relatively faster than the longer one for large number of isotropically
distributed collisions. 

However, the estimated timescale of this process seems to be much
longer than collisional lifetime of asteroids in the size range of 
several hundred meters to about $20\unit{km}$. Therefore, this effect
is probably not very important in formation of overall asteroid shapes
and their evolution, unless some of our assumptions is
unrealistic. For instance, if the exponent of the power-law projectile 
size distribution was smaller than $s=-2.574$, the erosion would be
faster. It is therefore important to set new observational constraints
on the small projectile population so that we can obtain more accurate
timescale estimates.

There is a group of newly discovered asteroids called Barbarians after its
first member, 234~Barbara, that are thought to be very old according
to their polarimetric and spectral properties
(\citealt{tanga2015}). These observations indicate rather high content
of calcium-aluminium-rich inclusions (\citealt{sunshine2008}). 
Collisional erosion may possibly play some role for larger long living
asteroids, but our present model cannot deal with reacumulated ejecta
effects, that are necessary to include for asteroids larger than about
$20\unit{km}$ in diameter. Further investigation of this area is
necessary.

\section*{Acknowledgments}
We are grateful to Alan Harris for his careful review that helped us
to improve the manuscript. TH acknowledges support from the project
RVO:67985815.

\bibliographystyle{mn2e}
\bibliography{elongation}

\begin{thebibliography}{}

\bibitem[\protect\citeauthoryear{{Asphaug}}{{Asphaug}}{2008}]{asphaug2008}
{Asphaug} E.,  2008, Meteoritics and Planetary Science, 43, 1075

\bibitem[\protect\citeauthoryear{{Bottke}, {Durda}, {Nesvorn{\'y}}, {Jedicke},
  {Morbidelli}, {Vokrouhlick{\'y}} \& {Levison}}{{Bottke}
  et~al.}{2005}]{bottke2005}
{Bottke} W.~F.,  {Durda} D.~D.,  {Nesvorn{\'y}} D.,  {Jedicke} R.,
  {Morbidelli} A.,  {Vokrouhlick{\'y}} D.,    {Levison} H.~F.,  2005, \icarus,
  179, 63

\bibitem[\protect\citeauthoryear{{Burchell} \&
  {Leliwa-Kopysty{\'n}ski}}{{Burchell} \&
  {Leliwa-Kopysty{\'n}ski}}{2010}]{burchell2010}
{Burchell} M.~J.,  {Leliwa-Kopysty{\'n}ski} J.,  2010, \icarus, 210, 707

\bibitem[\protect\citeauthoryear{{Catullo}, {Zappala}, {Farinella} \&
  {Paolicchi}}{{Catullo} et~al.}{1984}]{catullo1984}
{Catullo} V.,  {Zappala} V.,  {Farinella} P.,    {Paolicchi} P.,  1984, \aap,
  138, 464

\bibitem[\protect\citeauthoryear{{Cellino}, {Zappala} \& {Farinella}}{{Cellino}
  et~al.}{1991}]{cellino1991}
{Cellino} A.,  {Zappala} V.,    {Farinella} P.,  1991, \mnras, 253, 561

\bibitem[\protect\citeauthoryear{{Cibulkov{\'a}}, {Bro{\v z}} \&
  {Benavidez}}{{Cibulkov{\'a}} et~al.}{2014}]{cibulkova2014}
{Cibulkov{\'a}} H.,  {Bro{\v z}} M.,    {Benavidez} P.~G.,  2014, \icarus, 241,
  358

\bibitem[\protect\citeauthoryear{{Dohnanyi}}{{Dohnanyi}}{1969}]{dohnanyi1969}
{Dohnanyi} J.~S.,  1969, \jgr, 74, 2531

\bibitem[\protect\citeauthoryear{{Domokos}, {Sipos}, {Szab{\'o}} \&
  {V{\'a}rkonyi}}{{Domokos} et~al.}{2009}]{domokos2009}
{Domokos} G.,  {Sipos} A.~{\'A}.,  {Szab{\'o}} G.~M.,    {V{\'a}rkonyi} P.~L.,
  2009, \apjl, 699, L13

\bibitem[\protect\citeauthoryear{{Farinella}, {Vokrouhlick{\'y}} \&
  {Hartmann}}{{Farinella} et~al.}{1998}]{farinella1998}
{Farinella} P.,  {Vokrouhlick{\'y}} D.,    {Hartmann} W.~K.,  1998, \icarus,
  132, 378

\bibitem[\protect\citeauthoryear{{Farinella} \& {Zappal{\`a}}}{{Farinella} \&
  {Zappal{\`a}}}{1997}]{farinella1997}
{Farinella} P.,  {Zappal{\`a}} V.,  1997, Advances in Space Research, 19, 181

\bibitem[\protect\citeauthoryear{{Harris}}{{Harris}}{1979}]{harris1979b}
{Harris} A.~W.,  1979, \icarus, 40, 145

\bibitem[\protect\citeauthoryear{{Harris}}{{Harris}}{1990}]{harris1990}
{Harris} A.~W.,  1990, \icarus, 83, 183

\bibitem[\protect\citeauthoryear{{Harris} \& {Burns}}{{Harris} \&
  {Burns}}{1979}]{harris1979a}
{Harris} A.~W.,  {Burns} J.~A.,  1979, \icarus, 40, 115

\bibitem[\protect\citeauthoryear{{Harris}, {Fahnestock} \& {Pravec}}{{Harris}
  et~al.}{2009}]{harris2009}
{Harris} A.~W.,  {Fahnestock} E.~G.,    {Pravec} P.,  2009, \icarus, 199, 310

\bibitem[\protect\citeauthoryear{{Henych} \& {Pravec}}{{Henych} \&
  {Pravec}}{2013}]{henych2013}
{Henych} T.,  {Pravec} P.,  2013, \mnras, 432, 1623

\bibitem[\protect\citeauthoryear{{Holsapple}, {Giblin}, {Housen}, {Nakamura} \&
  {Ryan}}{{Holsapple} et~al.}{2002}]{holsapple2002}
{Holsapple} K.,  {Giblin} I.,  {Housen} K.,  {Nakamura} A.,    {Ryan} E.,
  2002, Asteroids III, pp 443--462

\bibitem[\protect\citeauthoryear{{Holsapple}}{{Holsapple}}{1993}]{holsapple1993}
{Holsapple} K.~A.,  1993, Annual Review of Earth and Planetary Sciences, 21,
  333

\bibitem[\protect\citeauthoryear{{Holsapple}}{{Holsapple}}{2003}]{holsapple-web}
{Holsapple} K.~A.,  2003, {Theory and equations for 'Craters from Impacts and
  Explosions'}, Available online:
  http://keith.aa.washington.edu/craterdata/scaling/\\theory.pdf, cited on 15
  April 2013.

\bibitem[\protect\citeauthoryear{{Holsapple} \& {Housen}}{{Holsapple} \&
  {Housen}}{2007}]{holsapple_housen2007}
{Holsapple} K.~A.,  {Housen} K.~R.,  2007, \icarus, 187, 345

\bibitem[\protect\citeauthoryear{{Housen}}{{Housen}}{2009}]{housen2009}
{Housen} K.,  2009, \planss, 57, 142

\bibitem[\protect\citeauthoryear{{Housen} \& {Holsapple}}{{Housen} \&
  {Holsapple}}{2011}]{housen2011}
{Housen} K.~R.,  {Holsapple} K.~A.,  2011, \icarus, 211, 856

\bibitem[\protect\citeauthoryear{{Kaasalainen}}{{Kaasalainen}}{2001}]{kaasalainen2001}
{Kaasalainen} M.,  2001, \aap, 376, 302

\bibitem[\protect\citeauthoryear{{Korycansky} \& {Asphaug}}{{Korycansky} \&
  {Asphaug}}{2001}]{korycansky2001}
{Korycansky} D.~G.,  {Asphaug} E.,  2001, in Lunar and Planetary Science
  Conference Vol.~32 of Lunar and Planetary Science Conference, {Shaping
  Asteroids via Small Impacts}.
p.~1433

\bibitem[\protect\citeauthoryear{{Korycansky} \& {Asphaug}}{{Korycansky} \&
  {Asphaug}}{2003}]{korycansky2003}
{Korycansky} D.~G.,  {Asphaug} E.,  2003, \icarus, 163, 374

\bibitem[\protect\citeauthoryear{{Melosh} \& {Ryan}}{{Melosh} \&
  {Ryan}}{1997}]{melosh1997}
{Melosh} H.~J.,  {Ryan} E.~V.,  1997, \icarus, 129, 562

\bibitem[\protect\citeauthoryear{{O'Brien} \& {Greenberg}}{{O'Brien} \&
  {Greenberg}}{2005}]{obrien2005}
{O'Brien} D.~P.,  {Greenberg} R.,  2005, \icarus, 178, 179

\bibitem[\protect\citeauthoryear{{Pravec} et~al.,}{{Pravec}
  et~al.}{2005}]{pravec2005}
{Pravec} P.,  et~al., 2005, \icarus, 173, 108

\bibitem[\protect\citeauthoryear{{Pravec} et~al.,}{{Pravec}
  et~al.}{2008}]{pravec2008}
{Pravec} P.,  et~al., 2008, \icarus, 197, 497

\bibitem[\protect\citeauthoryear{{Pravec} et~al.,}{{Pravec}
  et~al.}{2014}]{pravec2014}
{Pravec} P.,  et~al., 2014, \icarus, 233, 48

\bibitem[\protect\citeauthoryear{{Ronca} \& {Furlong}}{{Ronca} \&
  {Furlong}}{1979}]{ronca1979}
{Ronca} L.~B.,  {Furlong} R.~B.,  1979, Moon and Planets, 21, 409

\bibitem[\protect\citeauthoryear{{Stewart} \& {Leinhardt}}{{Stewart} \&
  {Leinhardt}}{2012}]{stewart2012}
{Stewart} S.~T.,  {Leinhardt} Z.~M.,  2012, \apj, 751, 32

\bibitem[\protect\citeauthoryear{{Sunshine}, {Connolly}, {McCoy}, {Bus} \& {La
  Croix}}{{Sunshine} et~al.}{2008}]{sunshine2008}
{Sunshine} J.~M.,  {Connolly} H.~C.,  {McCoy} T.~J.,  {Bus} S.~J.,    {La
  Croix} L.~M.,  2008, Science, 320, 514

\bibitem[\protect\citeauthoryear{{Szab{\'o}} \& {Kiss}}{{Szab{\'o}} \&
  {Kiss}}{2008}]{szabo2008}
{Szab{\'o}} G.~M.,  {Kiss} L.~L.,  2008, \icarus, 196, 135

\bibitem[\protect\citeauthoryear{{Tanga}, {Comito}, {Paolicchi}, {Hestroffer},
  {Cellino}, {Dell'Oro}, {Richardson}, {Walsh} \& {Delbo}}{{Tanga}
  et~al.}{2009}]{tanga2009}
{Tanga} P.,  {Comito} C.,  {Paolicchi} P.,  {Hestroffer} D.,  {Cellino} A.,
  {Dell'Oro} A.,  {Richardson} D.~C.,  {Walsh} K.~J.,    {Delbo} M.,  2009,
  \apjl, 706, L197

\bibitem[\protect\citeauthoryear{{Tanga} et~al.,}{{Tanga}
  et~al.}{2015}]{tanga2015}
{Tanga} P.,  et~al., 2015, \mnras, 448, 3382

\bibitem[\protect\citeauthoryear{{Thomas}}{{Thomas}}{1999}]{thomas1999}
{Thomas} P.~C.,  1999, \icarus, 142, 89

\bibitem[\protect\citeauthoryear{{Vokrouhlick{\'y}}, {Nesvorn{\'y}} \&
  {Bottke}}{{Vokrouhlick{\'y}} et~al.}{2003}]{vokrouhlicky2003}
{Vokrouhlick{\'y}} D.,  {Nesvorn{\'y}} D.,    {Bottke} W.~F.,  2003, \nat, 425,
  147

\bibitem[\protect\citeauthoryear{{Yanagisawa} \& {Hasegawa}}{{Yanagisawa} \&
  {Hasegawa}}{2000}]{yanagisawa2000}
{Yanagisawa} M.,  {Hasegawa} S.,  2000, \icarus, 146, 270

\bibitem[\protect\citeauthoryear{{Yanagisawa}, {Hasegawa} \&
  {Shirogane}}{{Yanagisawa} et~al.}{1996}]{yanagisawa1996}
{Yanagisawa} M.,  {Hasegawa} S.,    {Shirogane} N.,  1996, \icarus, 123, 192

\end{thebibliography}

\appendix
\section{Ejected mass timescale derivation}\label{app_time_mass}
Here we derive the relations for erosion timescale estimate based on
the excavated mass efficiency of projectiles from the
Sect.~\ref{ej_mass_mesure}. Eroded mass is given by the collision
probability of a target asteroid with a diameter $D_{\rm t}$ and a
projectile with a diameter $D_{\rm p}$, an efficiency of mass
excavation by the projectile given by its volume $V_{\rm p}$ and a
proportionality constant $A$ (it is given by the scaling laws of 
\citealt{holsapple-web}) per infinitesimal time interval $\d{}t$ times
projectile size distribution $n(D_{\rm p})$. The probability of
collision is given by the cross-sectional area of the sum of $D_{\rm
  t}$ and $D_{\rm p}$ times the intrinsic probability of collision
$p_{\rm i}$ (basically the flux of projectiles per area per time, see
\citealt{obrien2005})
\begin{equation}
 \d{}M_{\rm er}= \frac{1}{4}p_{\rm i}(D_{\rm t}+D_{\rm p})^2 AV_{\rm p}n(D_{\rm p})\d{}D_{\rm p}\d{}t\,.
 \label{app_diff_erod_mass}
\end{equation}
We plugin the projectile size distribution $n(D_{\rm p})=B D^s_{\rm
  p}$ (we used that from \citealt{bottke2005}), a spherical projectile
volume $V_{\rm p}=D^3_{\rm p}\upi/6$ and we integrate over a time
interval $\Delta t$, projectile size interval from $D_{\rm p,min}$ to
$D_{\rm p,max}$ on the right-hand side of
Eqn.~\ref{app_diff_erod_mass}. We also approximate $(D_{\rm 
  t}+D_{\rm p})^2\approx D^2_{\rm t}$ as the projectile's
diameter is much smaller than target's diameter (few percent for the
largest projectiles in our simulations). We can arbitrarily choose the
interval of integration of the left-hand side of
Eqn.~\ref{app_diff_erod_mass}. The reasonable choice is to put
$\int\!\!\d{}M_{\rm er}=0.5M_{\rm t,initial}$ because a catastrophic
disruption of an asteroid is defined the same way.

When we integrate Eqn.~\ref{app_diff_erod_mass}, we can finally evaluate
the time $\Delta t$ after which the target asteroid subjected to
consecutive collisions with a population of projectiles looses half
its original mass as
\begin{equation}
 \Delta t = \frac{12(s+4)M_{\rm t,initial}}{p_{\rm i}\upi ABD^2_{\rm
     t}(D_{\rm p,max\phantom{i}}^{s+4}\!\!-D_{\rm p,min}^{s+4})}\,.
 \label{app_timescale_mass}
\end{equation}
When we plugin $M_{\rm t,initial}\sim D_{\rm t}^3$ to
Eqn.~\ref{app_timescale_mass}, we see, that the timescale is
proportional to the target's mean diameter $D_{\rm t}$, inversely
proportional to $p_{\rm i}$ (smaller flux of projectiles leads to a
longer timescale) and inversely proportional to $A$, or erosion
efficiency of projectiles. 

\bsp

\label{lastpage}

\end{document}